\documentclass[journal]{IEEEtran}
\usepackage{graphicx}
\usepackage{amsmath}
\usepackage{subfig}

\begin{document}
\title{\mbox{A Layer Correlation Technique} for \mbox{Pion Energy Calibration} at
  the 2004 ATLAS \mbox{Combined Beam Test}}

\author{Karl-Johan Grahn,~\IEEEmembership{Member,~IEEE}\\
        On behalf of the ATLAS Liquid Argon Calorimeter Group
\thanks{Manuscript received November 13, 2009}
\thanks{K-J. Grahn is with the Royal Institute of Technology (KTH), Stockholm, Sweden (e-mail: kjg@particle.kth.se)}
}

\maketitle
\pagestyle{empty}
\thispagestyle{empty}

\begin{abstract}
A new method for calibrating the hadron response of a segmented
calorimeter is developed. It is based on a principal component
analysis of the calorimeter layer energy deposits, exploiting
longitudinal shower development information to improve the measured
energy resolution. Corrections for invisible hadronic energy and
energy lost in dead material in front of and between the ATLAS
calorimeters were calculated with simulated Geant4 Monte Carlo events
and used to reconstruct the energy of pions impinging on the
calorimeters during the 2004 Barrel Combined Beam Test at the CERN H8
area. For pion beams with energies between 20 and 180~GeV, the
particle energy is reconstructed within 3\% and the energy resolution
is improved by about 20\% compared to the electromagnetic scale.
\end{abstract}

\section{Introduction}

\begin{figure}[!t]
\centering
\includegraphics[width=2.4in, angle=270]{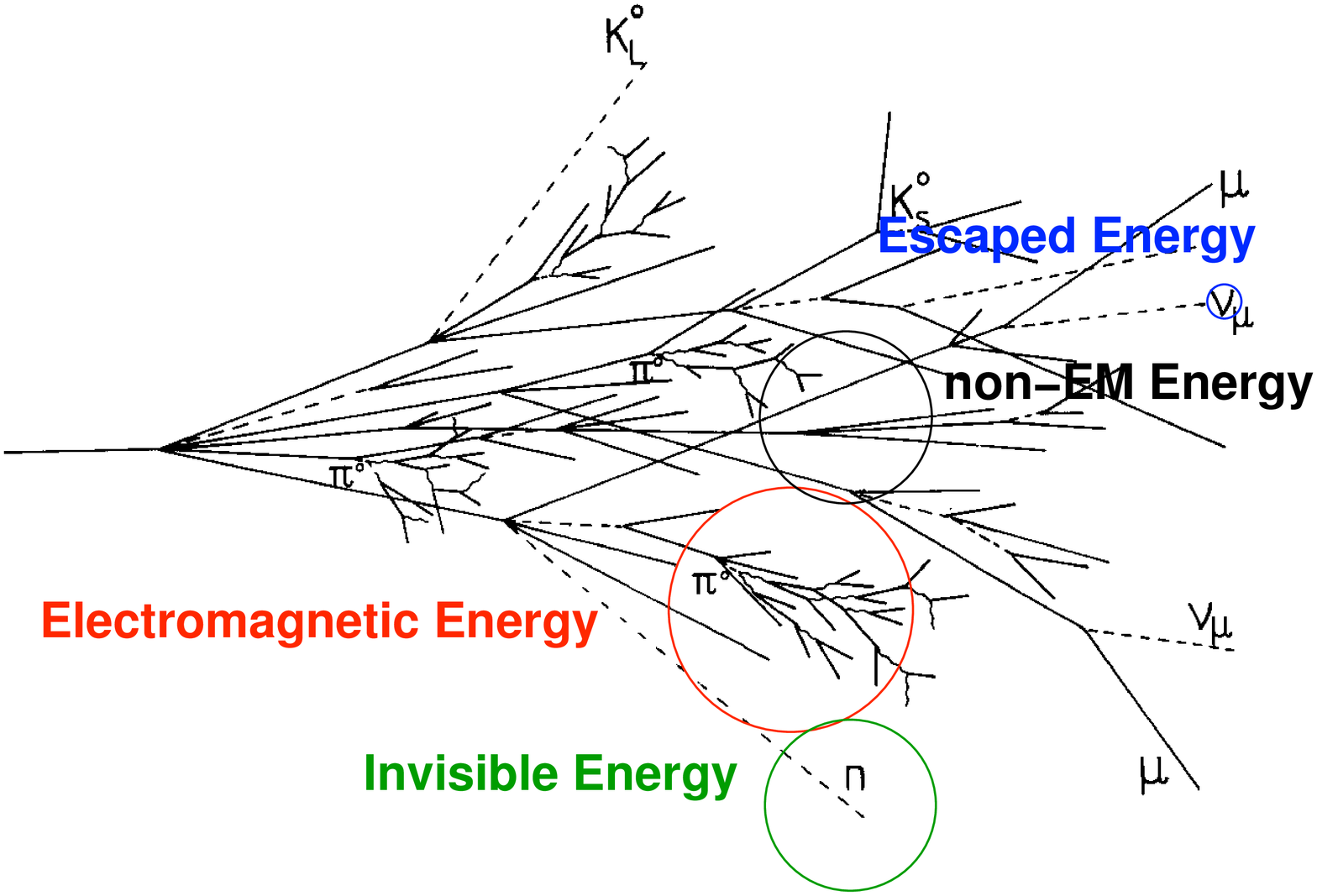}
\caption{Components of a hadronic shower. Based on figure
  in~\cite{grupen}.}
\label{hadshower}
\end{figure}

\IEEEPARstart{I}{n} general, the response of a calorimeter to hadrons
will be lower than for particles which only interact
electromagnetically, such as electrons and photons. This is due to
energy lost in hadronic showers in forms not measurable as an
ionization signal, i.e. nuclear break-up, spallation, and excitation,
energy deposits arriving out of the sensitive time window (such as
delayed photons), soft neutrons, and particles escaping the detector
(Fig. \ref{hadshower}). The shower has an electromagnetic and a
hadronic component. The size of the former increases with energy,
giving rise to a non-linear calorimeter response to impinging hadrons
if the calorimeter response to the electromagnetic and hadronic parts
of the shower is different. Such calorimeters are called
non-compensating. Moreover, hadronic showers exhibit large
even-by-event fluctuations, degrading the measured energy
resolution~\cite{wigmanscalo}.

The correlations between longitudinal energy deposits of the shower
have been shown~\cite{calocorr} to contain information on the
electromagnetic and hadronic nature of the shower. This calibration
method (called the Layer Correlation method in the following) aims to
use such information to improve energy resolution and linearity. It is
an alternative to the standard ATLAS hadronic calibration schemes.

\section{ATLAS and the 2004 Combined Beam Test}
ATLAS~\cite{detectorpaper} is one of the multi-purpose physics
experiments at the CERN Large Hadron Collider
(LHC)~\cite{lhcpaper}. Physics goals include searching for the Higgs
boson and looking for phenomena beyond the standard model of particle
physics, such as supersymmetry. Many measurements to be performed by
the LHC experiments rely on a correct reconstruction of hadronic final
state particles.

In the central barrel region, the ATLAS calorimeters consist of the
lead–-liquid argon (LAr) electromagnetic calorimeter and the
steel-–scintillator Tile hadronic calorimeter. Both are intrinsically
non-compensating.

\begin{figure}[!t]
\centering
\includegraphics[width=3.45in]{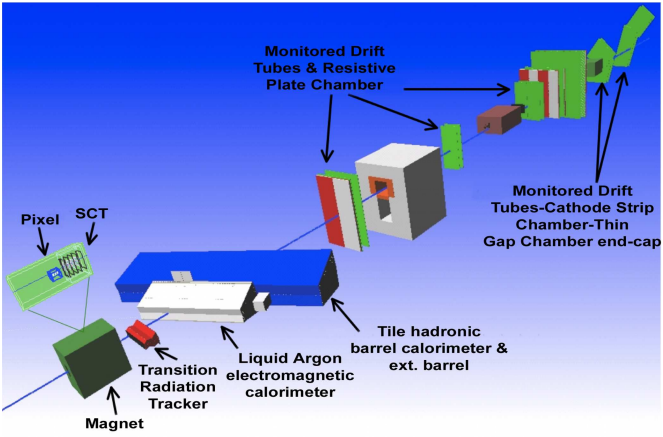}
\caption{2004 ATLAS Combined Beam Test set-up}
\label{cbt}
\end{figure}

The 2004 Combined Beam Test (Fig. \ref{cbt}) included a full slice of
the ATLAS Barrel region, including the pixel detector, the silicon
strip semiconductor tracker (SCT), the transition radiation tracker
(TRT), the LAr and Tile calorimeters and the muon spectrometer. In
addition, special beam-line detectors were installed to monitor the
beam position and reject background events. Those include beam
chambers monitoring the beam position and trigger scintillators. The
pixel and SCT detectors were surrounded by a magnet capable of
producing a field of 2~T, although no magnetic field was applied in
the runs used for this study.

The calorimeters were placed so that the beam impact angle
corresponded to a pseudorapidity\footnote{ATLAS has a coordinate
  system centered on the interaction point, with the $x$ axis pointing
  towards the center of the LHC ring, the $y$ axis pointing straight
  up, and the $z$ axis parallel to the beam. Pseudorapidity is defined
  as $-\ln(\tan(\theta/2))$, where $\theta$ is the angle to the
  positive $z$ axis.} of $\eta$ = 0.45 in the ATLAS detector. At this
angle, the expected amount of material in front of the calorimeters is
about 0.44~$\lambda_{I}$, where $\lambda_{I}$ is the nuclear
interaction length. This includes the LAr presampler. The LAr
calorimeter proper is longitudinally segmented in three layers that
extend in total to 1.3 $\lambda_{I}$. The dead material between the
Tile and LAr calorimeters spans about 0.6 $\lambda_{I}$. Finally the
three longitudinal segments of the Tile calorimeter stretch in total
for about 8.2 $\lambda_{I}$.

Events are selected by requiring signals in a trigger scintillator,
beam chambers, and the inner detector compatible with one particle
passing close to the nominal beam line. The TRT is used to reject
positrons by making a cut on the detected transition radiation.

The positive pion beam is known to have a sizable proton
contamination, which must be taken into account when deriving the
calibration, since the calorimeter response for pions and protons is
different. The fraction of protons in the beam was measured using the
differing probabilities of pions and protons to emit transition
radiation in the TRT. It was found to be 0\% at a beam energy of
20~GeV, 45\% at 50~GeV, 61\% at 100 GeV, and 76\% at 180~GeV.

\section{Method}
The calibration scheme consists of compensation weights and dead
material corrections. The former correct for the non-compensation of
the calorimeters, while the latter compensate for energy lost in
material with no calorimeter read-out.
\begin{align}
E_\textrm{corr}^\textrm{tot} = E_\textrm{tot}^\textrm{weighted} + E_\textrm{tot}^\textrm{DM}
\end{align}

The dead material corrections (see below) have an inherent dependence
on the beam energy. This dependence is removed by employing an
iteration scheme, where at each step the final estimated energy of the
former step is used, until the returned value is stable.

All corrections are extracted from a Geant4.7~\cite{geant1, geant2}
Monte Carlo simulation, with an accurate description of the Combined
Beam Test geometry. The QGSP\_BERT physics list was used. The
simulation gives access to both the true deposited energy in the
detector material, as well as the signal read out from the
calorimeters, including effects of the shaping electronics. The latter
is calibrated at the electromagnetic scale, i.e. giving the correct
deposited energy for electromagnetically showering particles, such as
electrons and photons. The corrections are calculated using a Monte
Carlo sample containing a scan of pion energies, from 15 to 230 GeV.

The energies of individual calorimeter cells are added up using a
topological cluster algorithm~\cite{topocluster}. The algorithm has
three adjustable thresholds: Seed ($S$), Neighbor ($N$), and Boundary
($B$). First, seed cells having an energy above the $S$ threshold are
found and a cluster is formed with this cell. Then, neighboring cells
having an energy above the $N$ threshold are added to the
cluster. This process is repeated until the cluster has no neighbors
with an energy above the $N$ threshold. Finally, all neighboring cells
having an energy above the $B$ threshold are added to the cluster. To
avoid bias, the absolute values of the cell energies are used. The
energy thresholds used for the $S$, $N$ and $B$ determination are set
to, respectively, four, two, and zero times the expected noise in a
given cell.

The reconstructed energy in a calorimeter layer $L$ is then obtained
by considering all the topological clusters in the event and summing
up the parts of the clusters that are part of that calorimeter layer.

\subsection{Eigenvectors of the covariance matrix}

In total there are seven longitudinal calorimeter layers (the LAr
presampler; the strips, middle, and back layers of the LAr
calorimeter; and the A, BC, and D layers of the Tile calorimeter). The
covariance matrix between these layers is calculated as
\begin{align}
\textrm{Cov}(M,L) = \left <E_{M}^\textrm{rec}E_{L}^\textrm{rec} \right> -
\left<E_{M}^\textrm{rec}\right>\left <E_{L}^\textrm{rec}\right>,
\label{eq:covmatrix}
\end{align}
where $M$ and $L$ denote calorimeter layers and $E_{M}^\textrm{rec}$
is the energy reconstructed at the electromagnetic scale in
calorimeter layer $M$.

An event can be regarded as a point in a seven-dimensional vector
space of calorimeter layer energy deposits. Its coordinates can be
expressed in a new basis of eigenvectors of the covariance
matrix. These eigenvectors are ordered by decreasing eigenvalue,
meaning that the projections along the first few eigenvectors contain
most of the information on event-by-event longitudinal shower
fluctuations. These projections are used as input to the calibration.

\begin{figure}[!t]
\centering
\includegraphics[height=0.45\linewidth, angle=270]{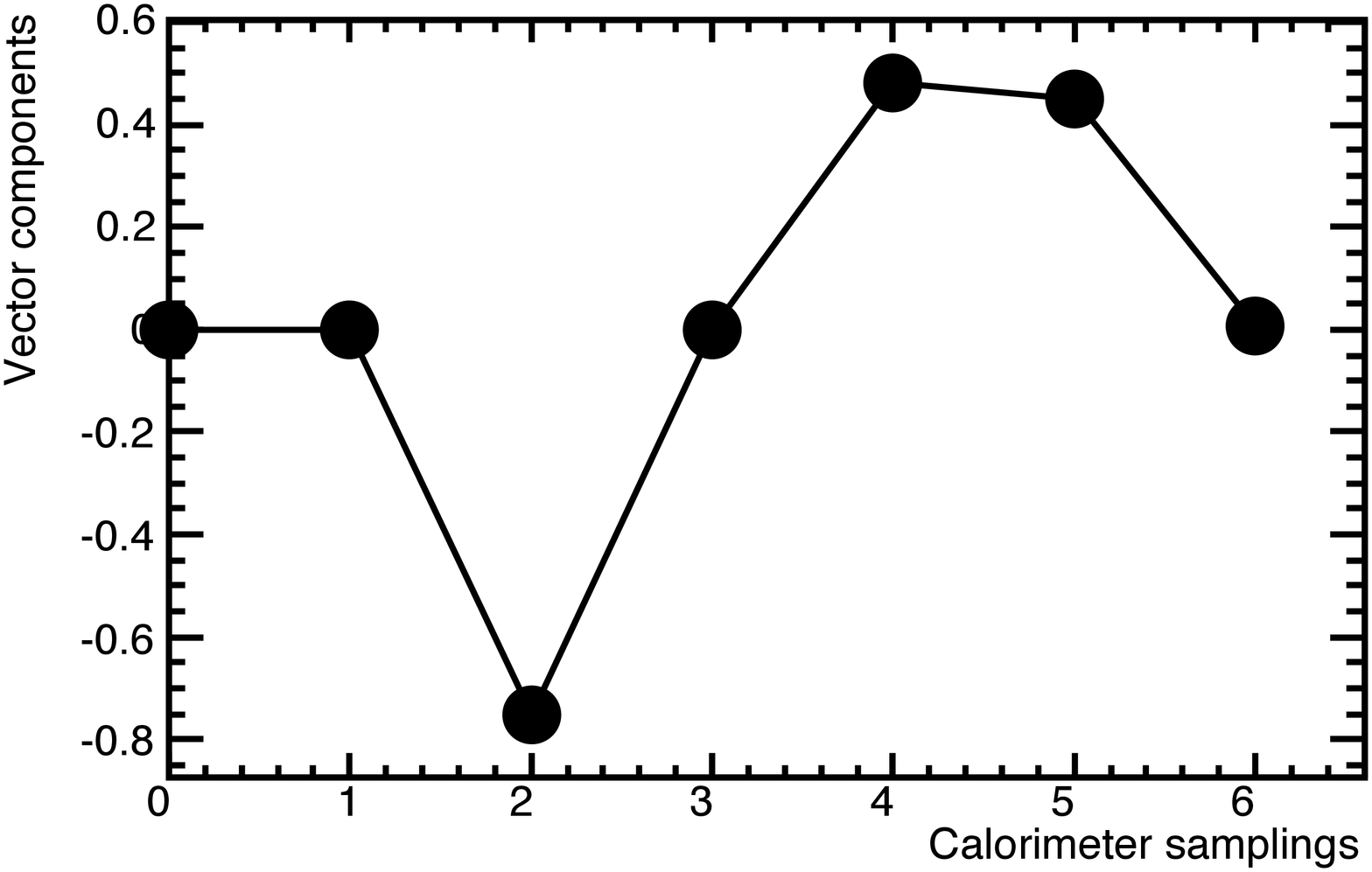}
\includegraphics[height=0.45\linewidth, angle=270]{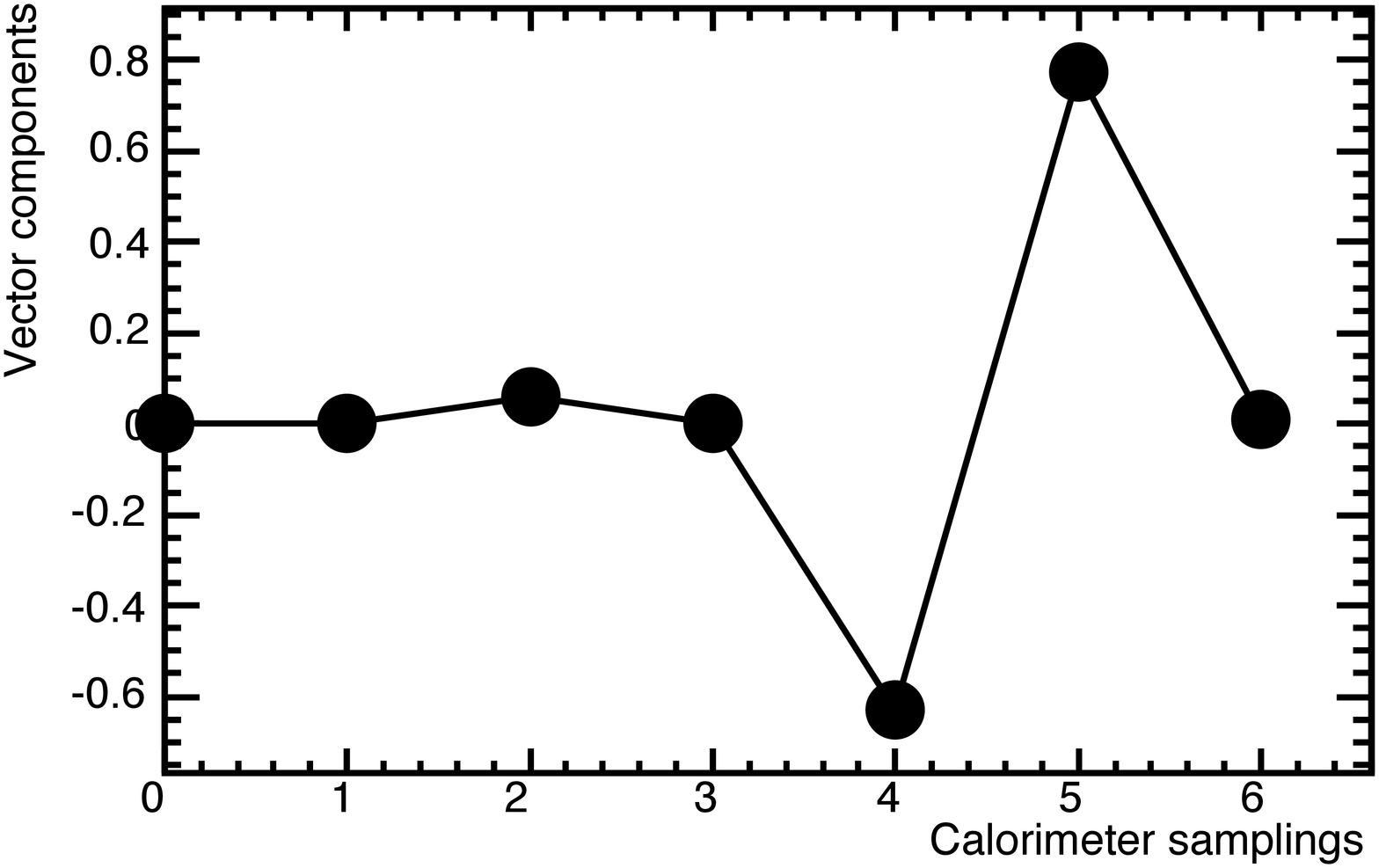}
\includegraphics[height=0.45\linewidth, angle=270]{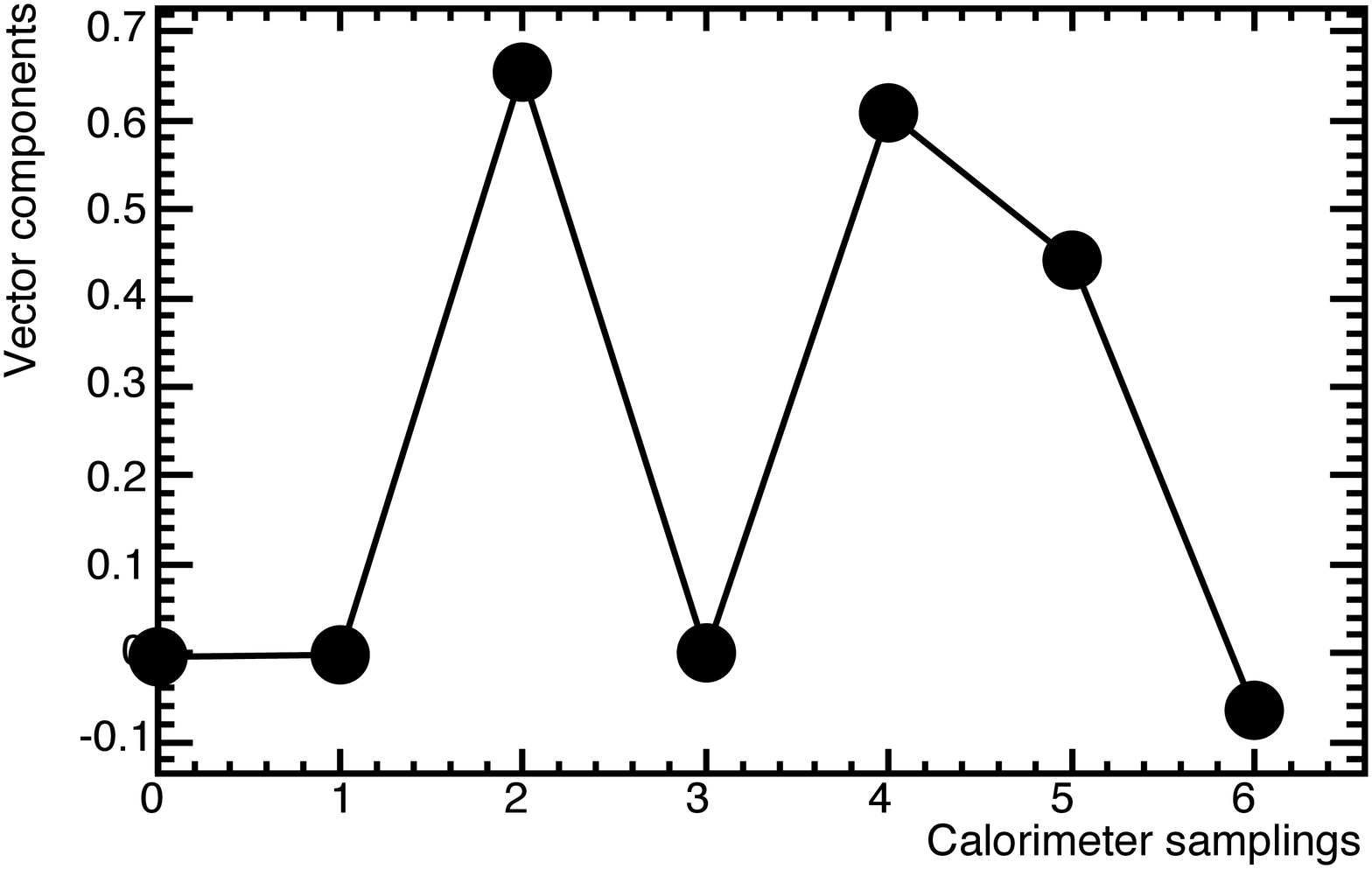}
\caption{The first three eigenvectors of the covariance matrix in the
  basis of the original calorimeter layers.}
\label{fig:eigvec}
\end{figure}

Fig. \ref{fig:eigvec} shows the first three eigenvectors of the
covariance matrix in the basis of the original calorimeter layers.

We find that
\begin{align*}
E^\textrm{rec}_\textrm{eig0} &\approx \frac{1}{\sqrt{6}} (-2 E_\textrm{LAr,middle} + E_\textrm{Tile,A} +E_\textrm{Tile,BC}),\\
E^\textrm{rec}_\textrm{eig1} &\approx \frac{1}{\sqrt{2}} (-E_\textrm{Tile,A}  + E_\textrm{Tile,BC}),\\
E^\textrm{rec}_\textrm{eig2} &\approx \frac{1}{\sqrt{3}} (E_\textrm{LAr,middle} + E_\textrm{Tile,A} +E_\textrm{Tile,BC}).
\end{align*}

Thus, the zeroth eigenvector is essentially a difference between the
Tile and LAr calorimeters, the first one a difference within the Tile
calorimeter and the second one a sum of both calorimeters. The rest of
the eigenvectors contain individual calorimeter layers.

\subsection{Compensation weights}

The compensation weights account for the non-linear response of the
calorimeters to hadrons. They are implemented as two-dimensional
128x128-bin lookup tables and are functions of the projections along
the first two (zeroth and first) eigenvectors of the covariance
matrix. Bi-linear interpolation is performed between the bins.

\begin{figure}[!t]
\centering
\includegraphics[width=\linewidth]{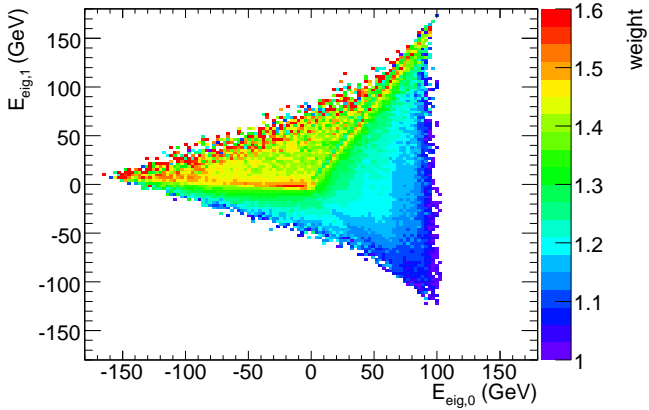}
\caption{Compensation weights for the first Tile calorimeter layer.}
\label{fig:weighttable}
\end{figure}

There is one weight table for each calorimeter layer, three for LAr
and three for Tile. The LAr presampler is not
weighted. Fig.~\ref{fig:weighttable} shows the weight table the first
Tile layer. The total reconstructed energy is the sum of the weighted
energies in each calorimeter layer.

\begin{align}
E_{L}^\textrm{weighted} &= w_{L}  E_{L}^\textrm{rec} \\
E_\textrm{tot}^\textrm{weighted} &=  \sum_{L} E_{L}^\textrm{weighted}
\end{align}

Fig. \ref{fig:weighttable} shows a weight table for the first layer of
the Tile calorimeter.

\subsection{Dead material correction}
\begin{figure}[!t]
\centering
\includegraphics[width=\linewidth]{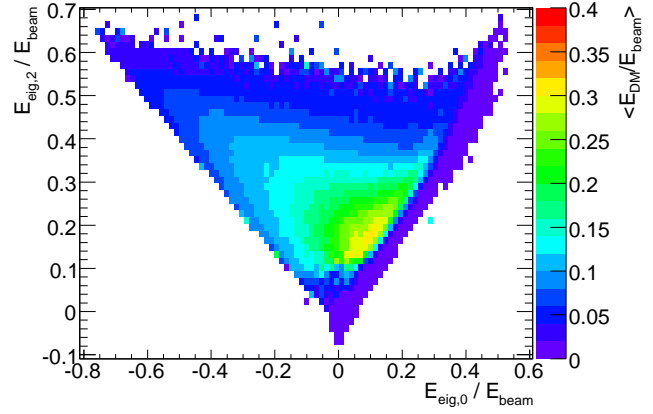}
\caption{Lookup table for dead material correction.}
\label{fig:dmcorr}
\end{figure}

Dead material is parts of the experiment that are neither active
calorimeter read-out material (liquid argon or scintillator), nor
sampling calorimeter absorbers (mostly lead or steel). Most of this
material is in the LAr cryostat wall between the LAr and Tile
calorimeters. There, pion showers are often fully developed, giving
rise to large energy loss. To correct for these losses, the
projections along the zeroth and second eigenvectors are used. When
making the lookup table both the eigenvector projections and the dead
material losses themselves are scaled with the true beam energy. Just
as for the compensation weights, the table (Fig. \ref{fig:dmcorr}) is
two-dimensional with 128x128 bins and bi-linear interpolation is
performed between the bins.

In addition, there is also dead material before the LAr calorimeter
(e.g. the inner detector) and leakage beyond the Tile
calorimeter. These losses are small in comparison to those between the
LAr and Tile calorimeters and were corrected for using a simple
parameterization as a function of beam energy.

\begin{equation}
E^\textrm{DM}_\textrm{other}(E_\textrm{beam}) = \left\{
 \begin{array}{rl}
  C_1 + C_2\,\sqrt{E_\textrm{beam}} & \textrm{if } E_\textrm{beam} < E_0 \\ 
  C_3 + C_4\,(E_\textrm{beam} - E_0) & \textrm{otherwise,}
 \end{array}\right.
\label{eq:otherdmcor}
\end{equation}
where $E_0 = 30000\ \mathrm{MeV}$. The resulting fitted parameters are
\begin{align}
C_1 & = (-75 \pm 31)\ \mathrm{MeV}\\
C_2 & = (5.78 \pm 0.22)\ \sqrt{\mathrm{MeV}}\\
C_3 & = (931 \pm 5)\ \mathrm{MeV}\\
C_4 & = 0.01435 \pm 0.0001
\end{align}

The final dead material correction is then the sum of these two contributions
\begin{equation}
E_\textrm{tot}^\textrm{DM} = E_\textrm{LArTile}^\textrm{DM} + E_\textrm{other}^\textrm{DM} 
\end{equation}

\section{Method validation}
Before applying it to beam test data, the calibration is validated on
a Monte Carlo sample statistically independent of the one used for
extracting the corrections. First, the performance of the compensation
weights is evaluated, then the linearity and resolution of the method
as a whole.

\subsection{Compensation}
\begin{figure}[!t]
\centering
\includegraphics[width=\linewidth]{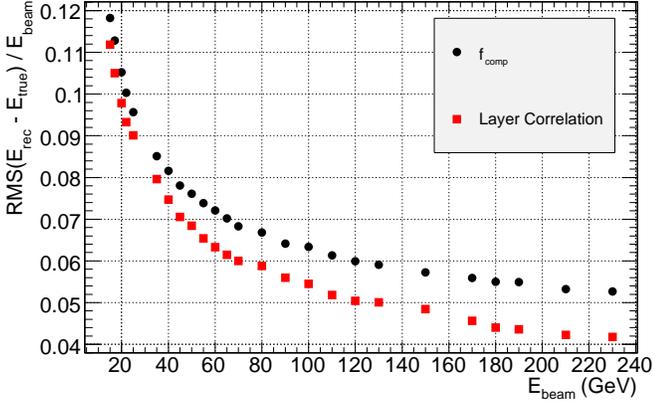}
\caption{Sample standard deviation of the difference between the
  weighted calorimeter energy and the true deposited energy as given
  by the simulation, as a function of beam energy.}
\label{fig:bias_rms}
\end{figure}

The resulting reconstructed energy after applying compensation weights
is compared to the true total deposited energy in the calorimeters as
given by the Monte Carlo simulation. The event-by-event difference
\begin{equation}
E_\textrm{tot}^{weighted} - E_\textrm{tot}^\textrm{true}(\textrm{calo})
\end{equation}
is considered. Fig.~\ref{fig:bias_rms} shows the sample standard
deviation of this variable as a function of beam energy. The
performance of the Layer Correlation technique is compared to that of
a simple calibration scheme where the energy of each event in the
sample is multiplied by a single factor ($f_\textrm{comp}$) calculated
to give the correct total deposited energy on average.

\subsection{Linearity and resolution}
\begin{figure*}[!t]
\centering
\includegraphics[width=6.17in]{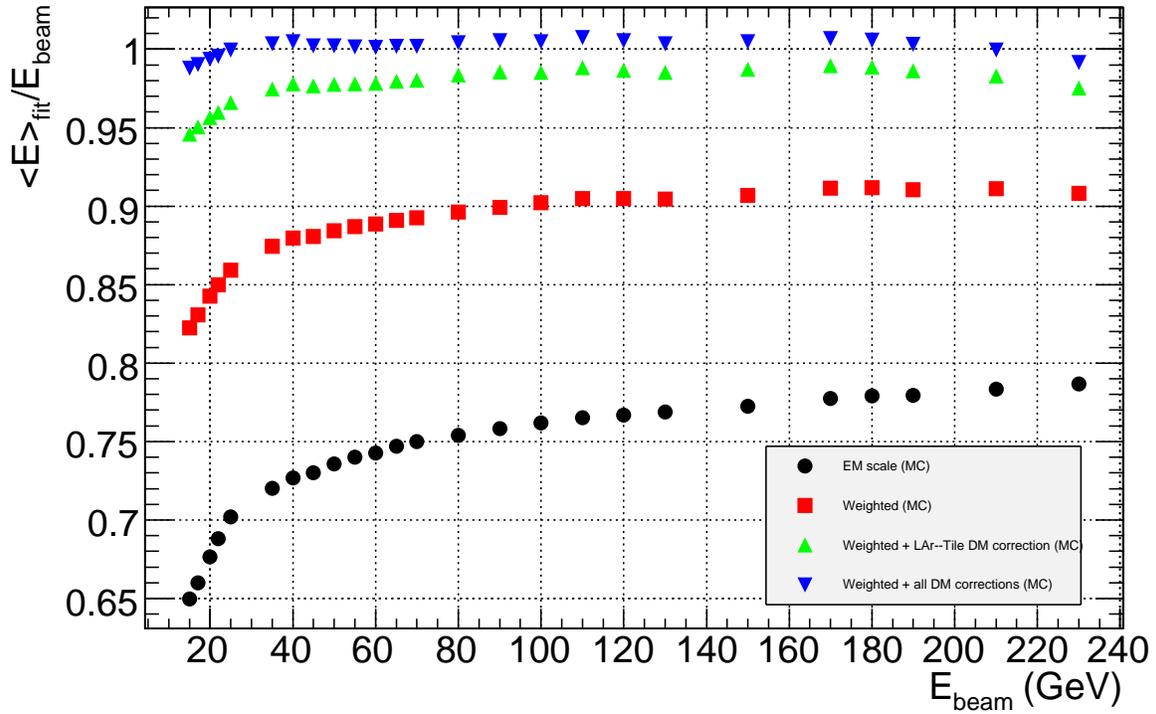}
\caption{Linearity of reconstructed energy as a function of beam
  energy when successively applying the different parts of the
  correction. Monte Carlo simulation with pions only.}
\label{fig:pion_lin}
\end{figure*}

\begin{figure*}[!t]
\centering
\includegraphics[width=6.17in]{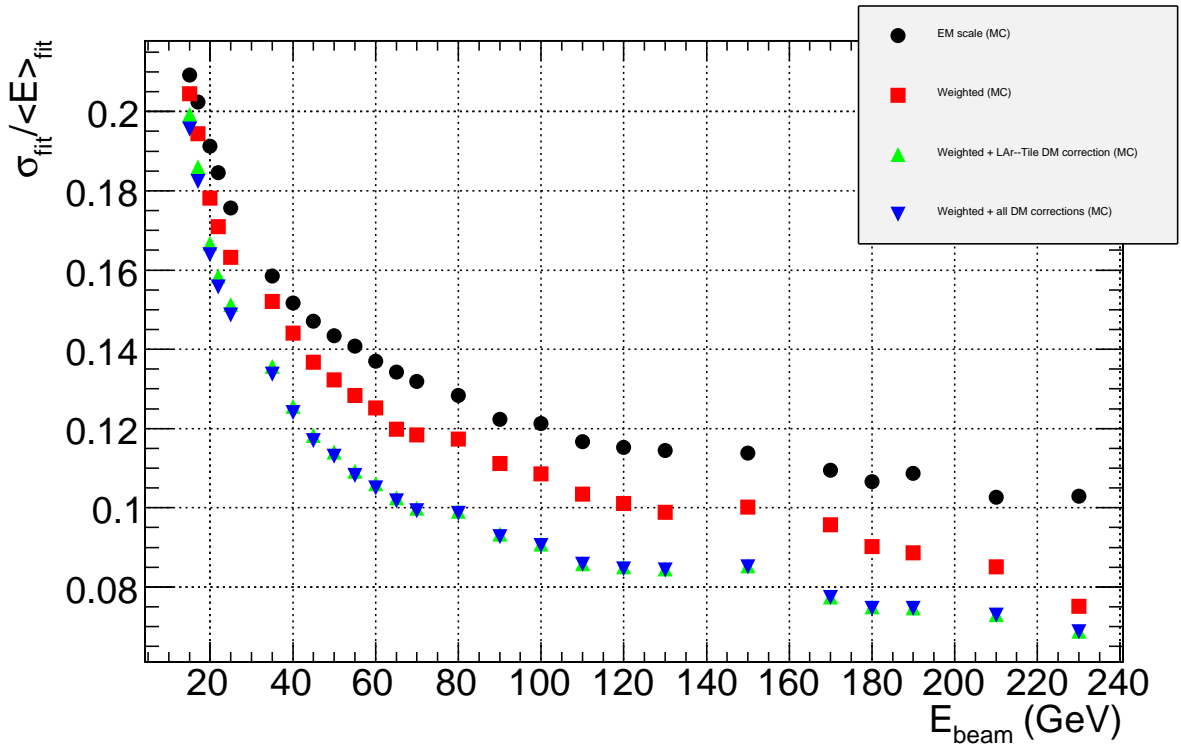}
\caption{Resolution of reconstructed energy as a function of beam
  energy when successively applying the different parts of the
  correction. Monte Carlo simulation with pions only.}
\label{fig:pion_res}
\end{figure*}

The performance for the fully corrected energy reconstruction is
assessed in terms of linearity (Fig.~\ref{fig:pion_lin}) and relative
resolution (Fig.~\ref{fig:pion_res}). The reconstructed energy
distribution is fitted with a Gaussian function in an interval of two
standard deviations on each side of the peak. This interval is found
iteratively. Linearity and resolution are shown -- first -- at the
electromagnetic scale -- then -- after successively applying the
corrections: compensation weights, the LAr--Tile dead material
correction, and finally after applying all corrections.

At the electromagnetic scale the calorimeter response is non-linear --
as expected -- and only about two thirds of the pion energy is
measured. Weighting recovers about 80\% to 90\% of the incoming pion
energy, while the LAr--Tile dead material correction accounts for an
additional 8\% to 10\%. After all corrections the pion energy is
correctly reconstructed within 1\% for all beam energies. Each
correction step makes the response more linear. The compensation
weights give the most important contribution to linearity improvement
at high energies, while the dead material effects play a more
significant role at low energies.

The relative resolution improves when applying each additional
correction step. At high beam energies (above 100 GeV) the
contribution of the compensation weights to the improvement in energy
resolution has the same magnitude as that of the LAr--Tile dead
material corrections, while at lower energies the dead material
corrections play a more important role.

\section{Application to Beam Test Data}
Finally, the method is applied to beam test data, which is compared
with Monte Carlo samples with a weighted mixture of pions and protons
to match the beam composition.

\begin{figure*}[!t]
\centering
\includegraphics[width=6.17in]{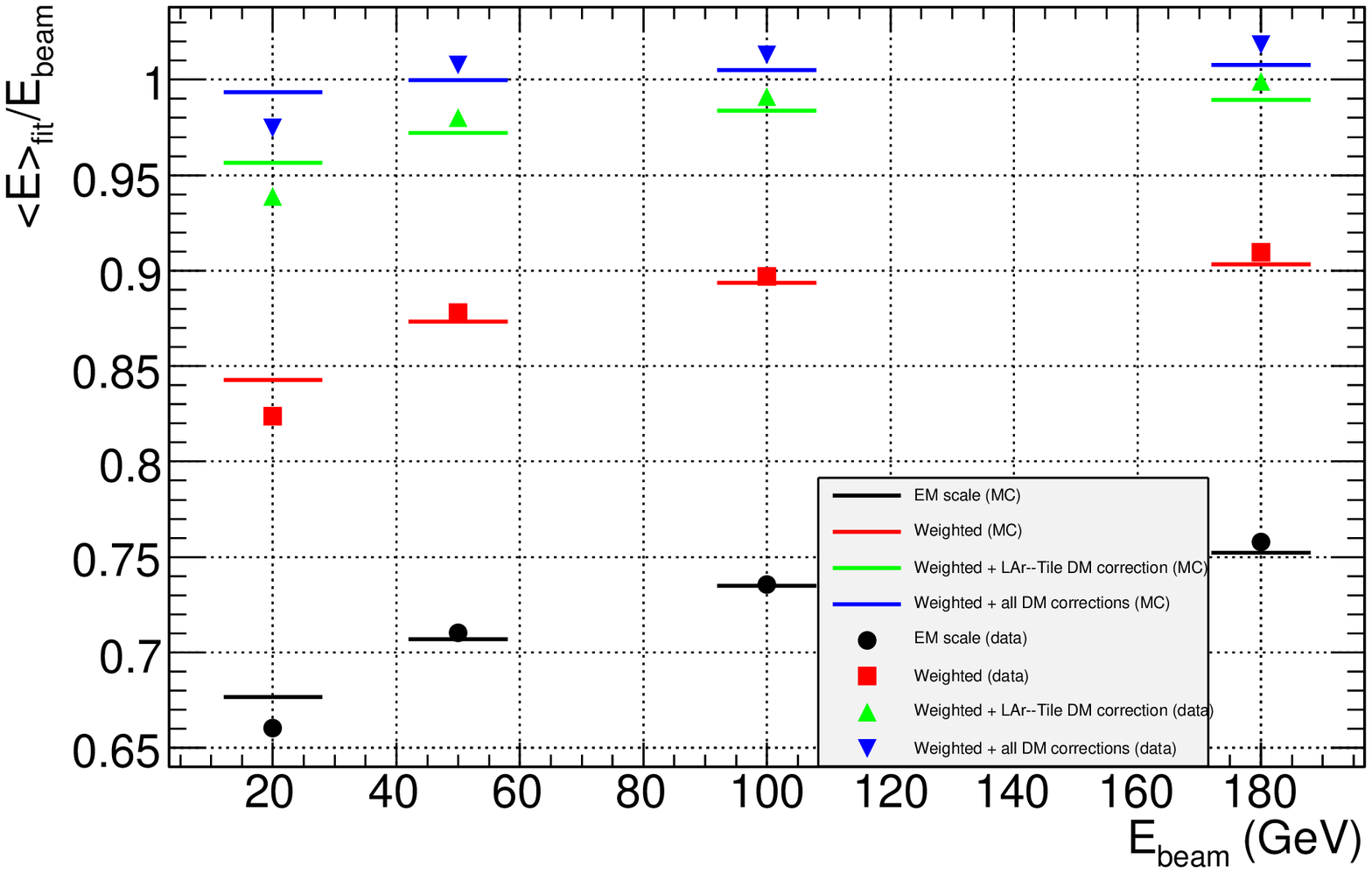}
\caption{Linearity of reconstructed energy as a function of beam
  energy when successively applying the different parts of the
  correction. Data (markers) and Monte Carlo simulation (horizontal
  lines) are shown. Mixed beam of pions and protons.}
\label{fig:data_lin}
\end{figure*}

\begin{figure*}[!t]
\centering
\includegraphics[width=6.17in]{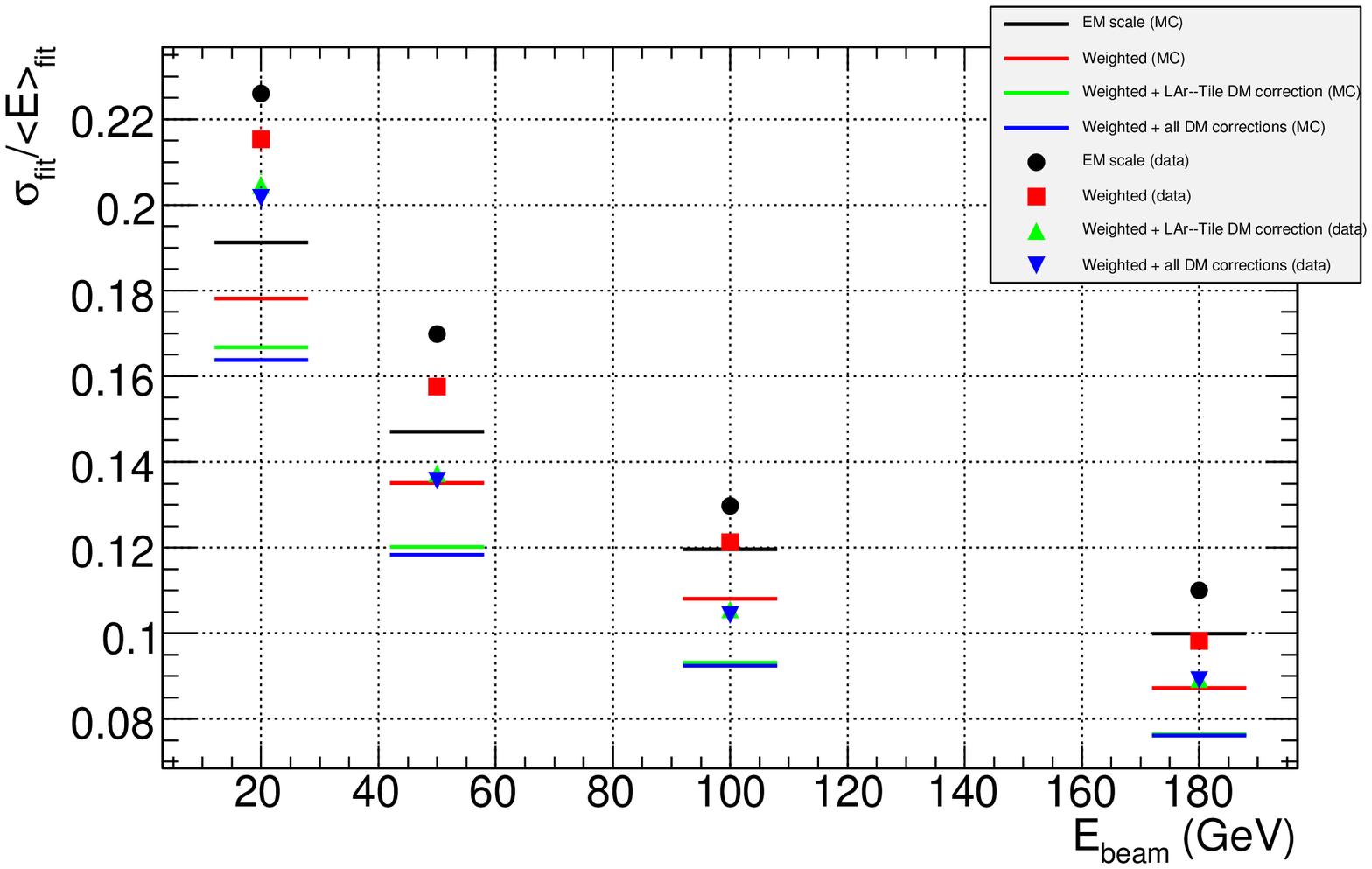}
\caption{Resolution of reconstructed energy as a function of beam
  energy when successively applying the different parts of the
  correction. Data (markers) and Monte Carlo simulation (horizontal
  lines) are shown. Mixed beam of pions and protons.}
\label{fig:data_res}
\end{figure*}

The linearity and relative resolution are shown in
Fig.~\ref{fig:data_lin} and Fig.~\ref{fig:data_res},
respectively. Again, the reconstructed energy is fitted with a
Gaussian function in an interval of two standard deviations on each
side of the peak.

After all corrections, linearity is recovered within 2\% for beam
energies above 50 GeV (3\% for 20 GeV).

The improvement in relative resolution when going from the
electromagnetic scale to applying all corrections is about 17\% to
22\% in data and 17\% to 29\% in simulation. The relative resolution
is smaller in Monte Carlo simulation than in data already at the
electromagnetic scale, by about 8 to 18\%, depending on beam
energy. When applying the corrections, the ratio of the relative
resolutions in data and simulation stays constant within 5\%.

\begin{figure}[!t]
\centering
\includegraphics[width=\linewidth]{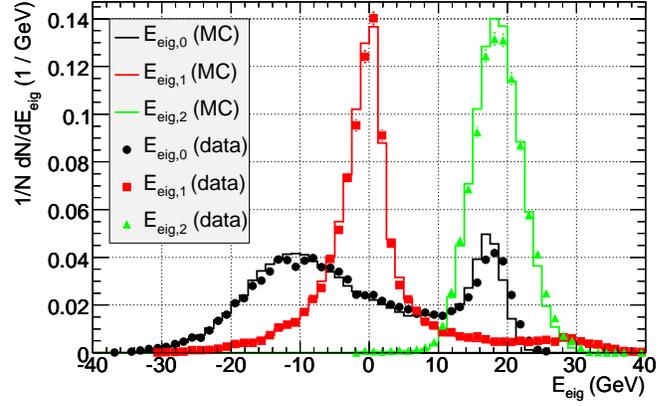}
\caption{Distribution of the first three eigenvector projections in data and Monte Carlo simulation.}
\label{fig:eigvecprojs}
\end{figure}

Fig.~\ref{fig:eigvecprojs} shows the distribution of the first three
eigenvector components for data and Monte Carlo simulation, with a
beam of 50 GeV particles. Good agreement is obtained between data and
simulation.

\begin{figure*}[!t]
\centering
\includegraphics[width=6in]{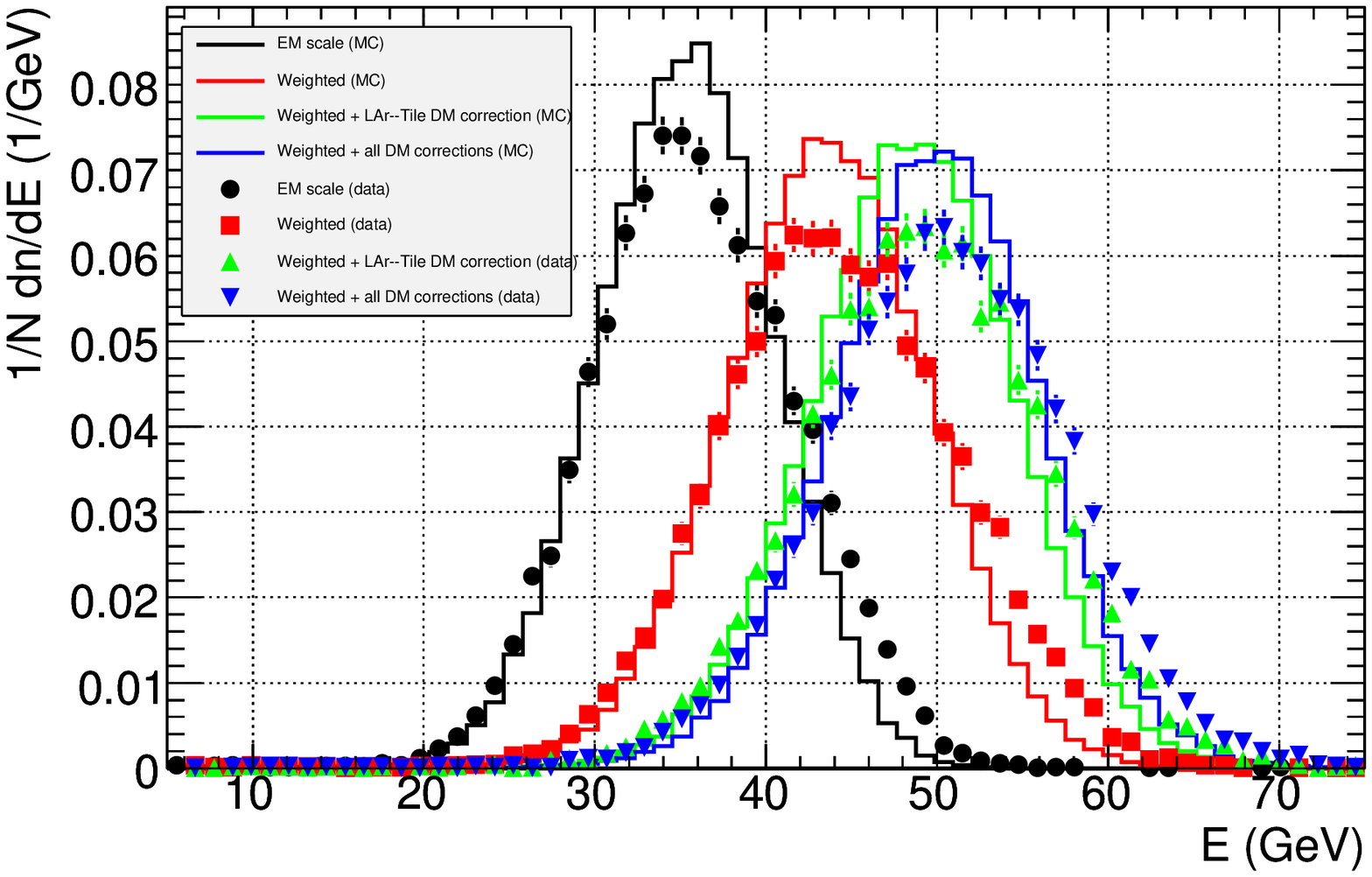}
\caption{Distribution of reconstructed energy at the different correction steps in data and Monte Carlo simulation.}
\label{fig:distr}
\end{figure*}

The shapes of the energy distributions for data and Monte Carlo
simulation for 50 GeV particles are compared in
Fig.~\ref{fig:distr}. The distribution in the Monte Carlo simulation
is narrower and less skewed than in the data. This is seen already at
the electromagnetic scale. The effect is even larger at 20 GeV but
less pronounced at higher energies.

\section{Conclusions}
The method was successfully applied to beam test data and is able to
reconstruct the incoming pion energy within 3\% in the energy range
20--180~GeV. Resolution is improved by about 20\% compared to the
electromagnetic scale.

The main deficiency of the Monte Carlo simulation is its inability to
correctly describe the energy resolution in the beam test
data. However, the relative improvement in resolution when applying
the calibration is similar in data and Monte Carlo simulation.

\appendices

\section*{Acknowledgment}
The essential contributions of Tancredi Carli, Francesco Span\`o, and
Peter Speckmayer are gratefully acknowledged.

\newpage
\bibliographystyle{IEEEtran}
\bibliography{record}

\end{document}